\documentclass[aps,prl,twocolumn,article,superscriptaddress,showpacs]{revtex4}
\usepackage{graphicx}
\usepackage{dcolumn}
\usepackage{bm}

\begin{document}

\title{Universal $\sqrt{2}\times\sqrt{2}$ structure and short-range charge order at the surfaces of BaFe$_{2-x}$Co$_{x}$As$_{2}$ compounds with various Co doping levels}

\author{Hui Zhang}
\affiliation{Hefei National Laboratory for Physical Sciences at Microscale and Department of Physics, University of Science and Technology of China, 96 JinZhai Road, Hefei, Anhui 230026, China
}
\author{Jun Dai}
\affiliation{Hefei National Laboratory for Physical Sciences at Microscale and Department of Physics, University of Science and Technology of China, 96 JinZhai Road, Hefei, Anhui 230026, China
}
\author{Yujing Zhang}
\affiliation{Hefei National Laboratory for Physical Sciences at Microscale and Department of Physics, University of Science and Technology of China, 96 JinZhai Road, Hefei, Anhui 230026, China
}
\author{Danru Qu}
\affiliation{Hefei National Laboratory for Physical Sciences at Microscale and Department of Physics, University of Science and Technology of China, 96 JinZhai Road, Hefei, Anhui 230026, China
}
\author{Huiwen Ji}
\affiliation{Hefei National Laboratory for Physical Sciences at Microscale and Department of Physics, University of Science and Technology of China, 96 JinZhai Road, Hefei, Anhui 230026, China
}
\author{G. Wu}
\affiliation{Hefei National Laboratory for Physical Sciences at Microscale and Department of Physics, University of Science and Technology of China, 96 JinZhai Road, Hefei, Anhui 230026, China
}
\author{X. F. Wang}
\affiliation{Hefei National Laboratory for Physical Sciences at Microscale and Department of Physics, University of Science and Technology of China, 96 JinZhai Road, Hefei, Anhui 230026, China
}
\author{X. H. Chen}
\affiliation{Hefei National Laboratory for Physical Sciences at Microscale and Department of Physics, University of Science and Technology of China, 96 JinZhai Road, Hefei, Anhui 230026, China
}
\author{Bing Wang}
\affiliation{Hefei National Laboratory for Physical Sciences at Microscale and Department of Physics, University of Science and Technology of China, 96 JinZhai Road, Hefei, Anhui 230026, China
}
\author{Changgan Zeng}
\thanks{cgzeng@ustc.edu.cn}
\affiliation{Hefei National Laboratory for Physical Sciences at Microscale and Department of Physics, University of Science and Technology of China, 96 JinZhai Road, Hefei, Anhui 230026, China
}
\author{Jinlong Yang}
\thanks{jlyang@ustc.edu.cn}
\affiliation{Hefei National Laboratory for Physical Sciences at Microscale and Department of Physics, University of Science and Technology of China, 96 JinZhai Road, Hefei, Anhui 230026, China
}
\author{J. G. Hou}
\affiliation{Hefei National Laboratory for Physical Sciences at Microscale and Department of Physics, University of Science and Technology of China, 96 JinZhai Road, Hefei, Anhui 230026, China
}

\date{\today}

\begin{abstract}
The structure and electronic order at the cleaved (001) surfaces of
the newly-discovered pnictide superconductors
BaFe$_{2-x}$Co$_{x}$As$_{2}$ with x ranging from 0 to 0.32 are
systematically investigated by scanning tunneling microscopy. A
$\sqrt{2}\times\sqrt{2}$ surface structure is revealed for all the
compounds, and is identified to be Ba layer with half Ba atoms
lifted-off by combination with theoretical simulation. A universal
short-range charge order is observed at this
$\sqrt{2}\times\sqrt{2}$ surface associated with an energy gap of
about 30 meV for all the compounds.

\end{abstract}


\maketitle


The newly discovered high-T$_{c}$ superconductors based on iron pnictides have ignited intense research interests, following the cuprate boom \cite{Kamihara,Wen,X. H. Chen,G. F. Chen,Ren,Rotter,Sefat}.
Although progresses have been made to understand their structural, electronic, and magnetic properties, the nature of superconductivity is yet a mystery \cite{Ishida}. Among the experimental techniques used to
investigate the high-T$_{c}$ superconductors, surface-sensitive
probes such as angle-resolved photoemission spectroscopy (ARPES) and scanning tunneling microscope (STM)
 have played a
critical role, due to their unrivaled energy and spatial/momentum
resolution. However the surface might differ from the bulk structurally and electronically, due to the symmetry-breaking and/or reconstruction. So thorough understanding of the surface structure and electronic properties at atomic scale is crucial to interpret correctly the experimental data based on surface-sensitive detections. On the other hand, the discrepant correlation between lattice, charge, and spin at the surface may lead to novel two-dimensional collective order, which would open a new door to the nanoelectronics applications based on correlated materials.

STM has been used to investigate the 122-type pnictide compounds, for which high-quality single crystals can be grown \cite{Rotter,Sefat}. Stripe \cite{Boyer,Massee1,Yin} or square-lattice \cite{Nascimento,Niestemski,Massee2} structures were observed at the surfaces of some certain compounds. However, the surface termination of the 122-type pnictides is still in debate, and the evolution of the geometric and electronic structures as a function of doping level has never been explored at atomic scale. In this letter, we report systematic STM study of electron-doped 122 compounds (BaFe$_{2-x}$Co$_{x}$As$_{2}$) with nominal doping composition of x = 0, 0.08, 0.17, 0.20, 0.25, 0.32.
A universal $\sqrt{2}\times\sqrt{2}$ surface structure is observed for
all the doping levels. With the aid of theoretical simulation, this
surface structure is determined to be well-ordered Ba plane with
half Ba atoms lifted-off. Strikingly, a short-range charge
order is characterized on such $\sqrt{2}\times\sqrt{2}$ surface for all the compounds
accompanied by an energy gap ranging from 24 to 40 meV.

The growth of pnictide single crystals and their magnetic/transport
characterization have been described elsewhere \cite{Wang}. The
crystals were cleaved in-situ at $\sim$120 K in ultra-high vacuum
environment with pressure lower than 2$\times$10$^{-10}$ mbar, and
immediately transferred into the STM stage, which were already
cooled to 5 K.

Two typical structures have been observed at the cleaved surfaces for all the investigated samples. The first is the widely observed stripe
structure (not shown) \cite{Boyer,Massee1,Yin}. The other type of surface is atomically flat, as shown in Fig. 1(a). The typical images with atomic resolution, for instance, for x = 0 and x = 0.20, are shown in Fig. 1(b) and (c), which have square lattice with lattice constant of $\sim$5.6 \AA \ . This observed universal square-like structure is quite striking, since
the structure is orthorhombic for x $<$ 0.20, while tetragonal for x
$>$ 0.20 at 5 K \cite{Rotter,Huang}. The typical STM images with
atomic resolution for x = 0 (parent) and x = 0.20 at 5 K are shown
in Fig. 1(b) and (c), and both show 5.6 \AA \ $\times$ 5.6 \AA \
lattice. Scanning tunneling spectroscopy (STS) measurement was also
performed on the cleaved surfaces with square-like lattice for all
the investigated samples at 5 K, and the typical dI/dV
spectroscopies taken away from defects are shown in Fig.
1(g). A protrusion feature at about -0.2 V is observed in the STSs
for all the samples, which could originate from collective ordering
or band-structure effect.

In the BaFe$_{2-x}$Co$_{x}$As$_{2}$ compounds consisting of FeAs layers
separated by a single Ba layer, the bond between Fe and As atoms is
quite strong, so the cleavage is likely to happen between the FeAs
layer and Ba layer, or in the Ba layer, leaving half of the Ba atoms
on each exposed surfaces \cite{Boyer}. The latter seems more
preferable energetically since this cleavage can still maintain
charge neutrality in the system. In order to determine the universal
surface structure, we did theoretical
calculation with half layer of Ba atoms at the surface, which
form $\sqrt{2}\times\sqrt{2}$ structure (in the tetragonal notation). Our calculations are carried out by employing the
Vienna \textit{ab initio} simulation package (VASP) using the
general gradient approximation (GGA) with PBE functional and plane
wave basis sets \cite{Kresse1,Kresse2,Blohl,Perdew}. A supercell geometry with 9 atomic layers
(4 FeAs layers and 5 Ba layers) and a vacuum layer about 30 \AA \
thick are used to simulate the surface. The plane wave cutoff is set
to be 600 eV. For geometry optimization, we use a
4$\times$4$\times$1 Monkhorst grid \cite{Monkhorst} to sample the
Brillouin zone, while a denser 6$\times$6$\times$1 grid is used for
density of states calculations. During the structural optimizations,
we fix the bottom 3 (2 Ba and 1 FeAs) layers in the bulk
configuration and allow all other atoms in the supercell to move
until all forces vanished within 0.02 eV/\AA \ .

The simulated STM images for the parent with orthorhombic and
tetragonal structures are displayed in Fig. 1(d), and both show
square-like pattern. No essential difference can be resolved from
our simulation, although the symmetries of the two structures
diverge. It is reasonable considering that the distortion of the
low-T orthormhombic structure from the high-T tetragonal structure
in BaFe$_{2}$As$_{2}$ is quite minor \cite{Rotter,Huang}. The simulated
STM image for x $=$ 0.25 (tetragonal structure) is shown in Fig.
1(e), which is similar to that of the parent compound. Orthorhombic
distortion also has little effect on the simulated images. So the
simulated STM images with half Ba layer termination well reproduce
the experimental data for both undoped and doped compounds. STM
simulation for the parent compound with FeAs layer as the surface
termination was also performed, and only shows a square-like lattice
of 4 \AA \ $\times$ 4 \AA \ (or 1 $\times$ 1 in the tetragonal
notation). This is inconsistent with the observed STM images.

\begin{figure}

\includegraphics[width=0.5\textwidth]{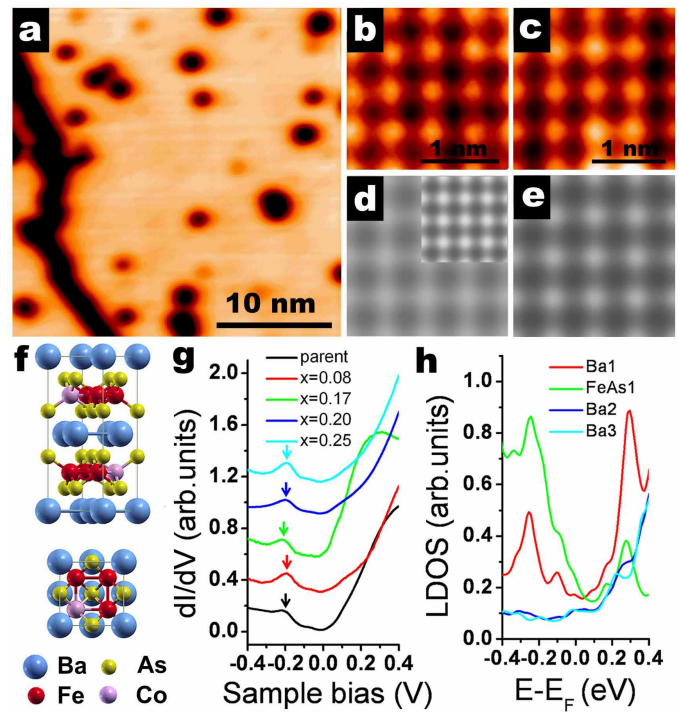}

\caption{(a) Large-scale STM image of the
BaFe$_{1.8}$Co$_{0.2}$As$_{2}$ surface with $\sqrt{2}\times\sqrt{2}$ structure
(V$_{s}$ $=$ 0.5 V, I $=$ 0.5 nA). (b) and (c) are STM
images showing the $\sqrt{2}\times\sqrt{2}$ lattice with atomic
resolution on the surfaces of BaFe$_{2}$As$_{2}$ (V$_{s}$ = 30 meV, I
= 3 nA) and BaFe$_{1.8}$Co$_{0.2}$A$s_{2}$ (V$_{s}$ = 2 meV, I = 20 nA),
respectively. (d) and (e) are the simulated empty-state STM images
with half Ba atoms on top of BaFe$_{2}$As$_{2}$ and
BaFe$_{1.75}$Co$_{0.25}$As$_{2}$, respectively. (f) The side view and
top view of the calculated structure of
BaFe$_{1.75}$Co$_{0.25}$As$_{2}$. The calculated structure is
orthorhombic in (d), and tetragonal in (e) and the inset of (d). (g)
dI/dV spectroscopies for BaFe$_{2-x}$Co$_{x}$As$_{2}$ with x $=$ 0,
0.08, 0.17, 0.20, and 0.25, respectively. (h) Calculated LDOS for
different layers in BaFe$_{2}$As$_{2}$.}
\end{figure}

We also calculated the local density of states (LDOS) of the parent
for different layers, and there is indeed a feature arising around
-0.2 eV for the first Ba and FeAs layer, which is consistent with
the STS results. The agreement between experimental STM/STS results
and theoretic simulations strongly suggests that the cleaved surface
for the series of BaFe$_{2-x}$Co$_{x}$As$_{2}$ compounds with x ranging
from 0 to 0.32 is terminated by half layer of Ba atoms. The half Ba
atoms form well-ordered $\sqrt{2}\times\sqrt{2}$ structure, which is
insensitive to the orthorhombic structural distortion.

STS in narrower range around Fermi level was measured on the
$\sqrt{2}\times\sqrt{2}$ surface for all the investigated samples,
and the typical spatially resolved dI/dV spectroscopies obtained
away from defects for x $=$ 0, 0.20, and
0.32 are shown in Fig. 2. The dI/dV curves on all the samples show similar shapes, and
surprisingly a weak gap structure is observed for all the
samples. The gap size 2$\Delta$ is similar for all the samples,
ranging from 24 to 40 meV. This gap size is much larger than those
previously observed by STM ($\sim$12 meV) \cite{Massee1,Yin} and
ARPES (13 and 10 meV for different Fermi surfaces) \cite{Terashima}
on the striped surface, which were attributed to the isotropic
s-wave superconducting gap. The universality of the observed gap
feature on the $\sqrt{2}\times\sqrt{2}$ surfaces strongly indicates
that they share the same origin. Superconducting gap due to bulk
superconductivity is excluded since the gap was observed in the
parent compound. Neither can the band structure of the $\sqrt{2}\times\sqrt{2}$ surface termination give rise to such a gap from our
calculation. It also can not be attributed to the bulk SDW order, as
SDW disappears when x $>$ 0.20 \cite{Wang}. It could be the
manifestation of some surface collective order induced by the new $\sqrt{2}\times\sqrt{2}$ surface termination, such as surface
superconductivity, surface spin order or surface charge order.
Occasionally gap feature becomes very weak at some random
locations (blue curves in Fig. 2). The observation of universal gap
feature indicates that the electronic correlation on the surfaces
depart significantly from the bulk, and that the prediction of bulk
properties based on surface-sensitive probes should be very careful.

\begin{figure}
\begin{center}
\includegraphics[width=0.5\textwidth]{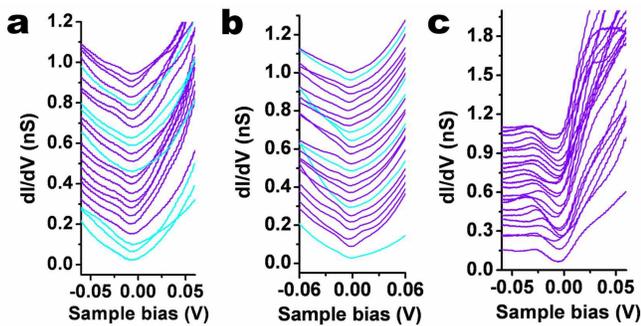}
\end{center}
\caption{DI/dV spectroscopies measured along line trajectories on
the BaFe$_{2-x}$Co$_{x}$As$_{2}$ compounds with x $=$ 0 (a), 0.20 (b),
and 0.32 (c). The lengths of the trajectories are 31 nm, 32 nm, and
28 nm, respectively. The purple and blue curves represent
spectroscopies with and without gap feature, respectively.}
\end{figure}

In order to explore possible charge ordering on the surface, dI/dV
mapping simultaneously with topography imaging was performed on all
the investigated samples at different bias voltage. Typical STM
images and the corresponding dI/dV maps for x $=$ 0.17 are shown in
Fig. 3(a) and (f) with sample bias of 0.02 V and -0.02 V,
respectively. The $\sqrt{2}\times\sqrt{2}$ lattice with atomic
resolution is clearly resolved in both topography and dI/dV images
at such low tunneling-junction resistance. Furthermore,
inhomogeneous corrugations at nanoscale are observed, superimposed
on the square lattice. This is also evidenced in Fig. 1(b) and (c).
If the tunneling current is lowered while the sample bias is unchanged, the atomic resolution becomes
weaker as shown in Fig. 3(b) and (g), due to the increased distance
between the tip and the substrate, while the nanoscale corrugations
are still preserved. When the bias voltage is swept, the spatial
inhomogeneity pattern in the same area also varies correspondingly,
and even reverses at some locations as clearly shown in Fig. 3. So
the nanoscale inhomogeneity is unlikely to originate from
topographic effect, and should be due to electronic effect.

\begin{figure}
\begin{center}
\includegraphics[width=0.5\textwidth]{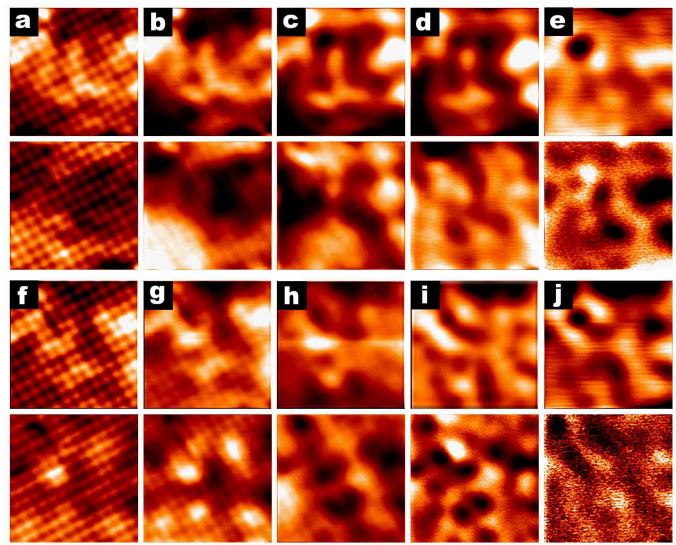}
\end{center}
\caption{A series of topography images (upper) and corresponding dI/dV
maps (lower) measured at the same area with the size of 6.25 nm
$\times$ 6.25 nm when varying the tunneling parameters: (a) V$_{s}$
$=$ 0.02 V, I $=$ 2 nA; (b) V$_{s}$ $=$ 0.02 V, I $=$ 0.5 nA; (c)
V$_{s}$ $=$ 0.05 V, I $=$ 0.5 nA; (d) V$_{s}$ $=$ 0.1 V, I $=$ 0.5
nA; (e) V$_{s}$ $=$ 0.4 V, I $=$ 0.5 nA; (f) V$_{s}$ $=$ -0.02 V, I
$=$ 1.5 nA; (g) V$_{s}$ $=$ -0.02 V, I $=$ 0.5 nA; (h) V$_{s}$ $=$
-0.05 V, I $=$ 0.5 nA; (i) V$_{s}$ $=$ -0.1 V, I $=$ 0.5 nA; (j)
V$_{s}$ $=$ -0.4 V, I $=$ 0.5 nA.}
\end{figure}

The larger-scale STM images and corresponding dI/dV maps for
different doping concentrations are shown in Fig. 4. It is clear
that the inhomogeneous charge order is a universal phenomenon, which
can be observed for all the compounds, including the parent compound,
although it is relatively weak. The inhomogeneity pattern does not
have spatial long-range order as indicated by the fast Fourier
transformation (not shown), nor obvious preferable orientation. This
short-range charge order resembles the spatial charge fluctuation in
YSi$_{2}$ nanowires \cite{Zeng}. The evolution of the charge
inhomogeneity as a function of the sample bias is also investigated.
It is interesting that the spatial pattern and characteristic length
of the charge inhomogeneity vary when the voltage is changed. We did
autocorrelation calculation for the dI/dV images for quantitative
analysis. The correlation lengths can be extracted from the radical
line profiles, which characterize the mean distance between the
neighboring maxima/mimima \cite{Bubendorff}. The
correlation length depending on the doping level (x $=$ 0.17, 0.20,
and 0.25) and energy is shown in Fig. 4(i). Surprisingly, the
correlation length at a given energy and its evolution with energy
is quite similar for different doping levels, although they show different order and/or ordering strength in the bulk. The
averaged correlation length is about 3.0 nm. Since the charge inhomogeneity and energy gap
are universally observed on the $\sqrt{2}\times\sqrt{2}$ surface for
all the investigated compounds with a wide doping range, it is
natural to think that they are connected, i.e., the energy gap
reflects the characteristic energy of the short-range charge order.

\begin{figure}
\begin{center}
\includegraphics[width=0.5\textwidth]{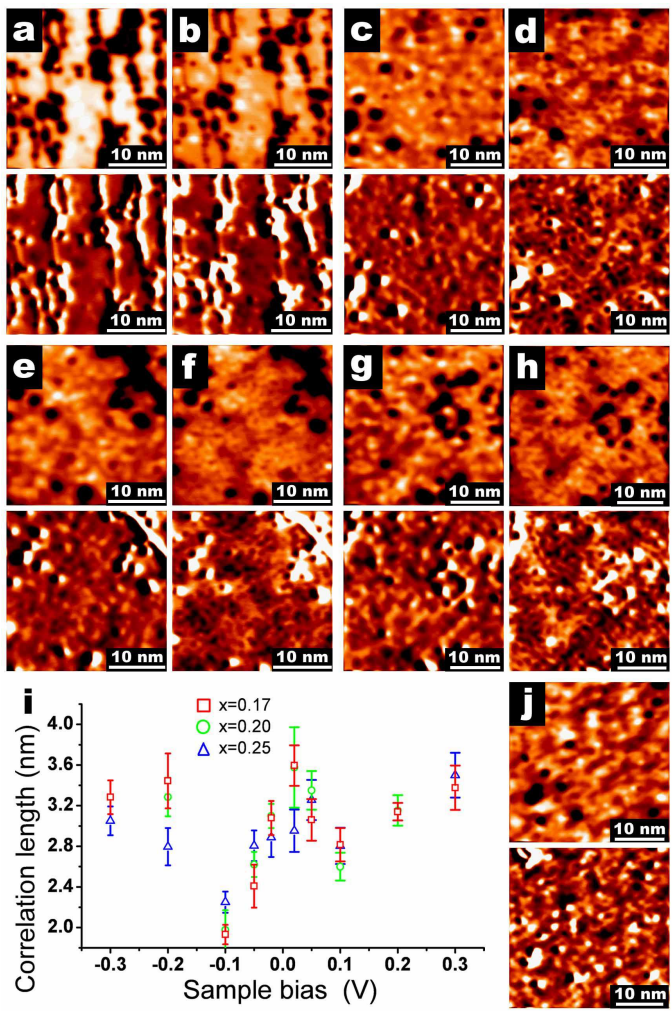}
\end{center}
\caption{Large-scale topography (upper) and dI/dV (lower) images of
BaFe$_{2-x}$Co$_{x}$As$_{2}$ for x $=$ 0 (a, b), 0.17 (c, d), 0.20 (e,
f), 0.25 (g, h). The sample bias is 0.1 V for (a), (c), (e), (g) and
-0.1 V for (b), (d), (f), (h). The tunneling current is 0.5 nA for
all the images. The images for the same doping concentration are in
the same area. (i) Correlation length derived from the dI/dV images as a
function of sample bias for x $=$ 0.17, 0.20 and 0.25. (j) Topography and dI/dV images for x $=$ 0.17 at V$_{s}$ of -0.01 V.  }
\end{figure}

For cuprates, spatially periodic charge modulations have been seen
within the energy gap  due to quasiparticle interference oscillations
in the superconducting state \cite{McElroy} or electronic
modulations in the pseudogap state \cite{Vershinin,Hanaguri}. The
periodicity is energy dependent for the former and independent for
the latter. We acquired topography and dI/dV images with sample bias
of -0.01 V, within the energy gap as displayed in Fig. 4(j). The
observed charge pattern does not show any regular features. This
might reflect the different pairing symmetry of pnictides (extended s-wave probably \cite{Mazin}) from cuprates (d-wave).

DFT calculation failed to reproduce such charge inhomogeneity in
both the parent and doped compounds as shown in Fig. 1(b) and (c).
It has been predicted theoretically that electron interactions in
iron pnictides are sufficiently strong to produce incipient Mott
physics \cite{Si}. The charge order might come from the enhanced
electron correlation effect at the symmetry-broken
$\sqrt{2}\times\sqrt{2}$ surface, though more efforts are needed to
identify the microscopic origin.

In conclusion, a $\sqrt{2}\times\sqrt{2}$ surface structure
terminated by half layer of Ba atoms is identified at the surfaces
of BaFe$_{2-x}$Co$_{x}$As$_{2}$ compound with x varying from 0 to
0.32, which is insensitive to the orthorhombic distortion. A
universal short-range charge order accompanied by an energy gap of
around 30 meV appears at this $\sqrt{2}\times\sqrt{2}$ surface for
all the investigated Co doping levels. This universal charge order
suggests relatively strong electron correlation in the iron
pnictides, at least on the symmetry-broken surfaces. It also sheds
light on the exploration of novel cooperative effects of the
correlated materials in reduced dimensions.

We thank Zhenyu Zhang, Hanno H. Weitering, Dimitrie Culcer, and Donglai Feng for helpful discussions. This work was supported by NKBRPC (2009CB929504, 2006CB922004), NSFC (50532040, 10825415), NCET, and CAS.

\end{document}